\documentclass[aps,prd,twocolumn,superscriptaddress,showpacs]{revtex4}

\usepackage{epsfig}
\usepackage{graphicx}

\usepackage{dcolumn}
\usepackage{bm}

\begin{document}

\title{Limits on spin-dependent WIMP-nucleon cross-sections from \\ the XENON10 experiment}

\author{J. Angle} \affiliation{Department of Physics, University of Florida, Gainesville, FL 32611, USA} 
\affiliation{Physics Institute, University of Z\"urich, Z\"urich, CH-8057, Switzerland}
\author{E. Aprile} \affiliation{Department of Physics, Columbia University, New York, NY 10027, USA}
\author{F. Arneodo} \affiliation{INFN - Laboratori Nazionali del Gran Sasso, Assergi, 67100, Italy}
\author{L. Baudis} 
\affiliation{Physics Institute, University of Z\"urich, Z\"urich, CH-8057, Switzerland}
\author{A. Bernstein} \affiliation{Lawrence Livermore National Laboratory, 7000 East Ave., Livermore, CA 94550, USA}
\author{A. Bolozdynya} \affiliation{Department of Physics, Case Western Reserve University, Cleveland, OH 44106, USA}
\author{L.C.C. Coelho} \affiliation{Department of Physics, University of Coimbra, Coimbra 3004-516, Portugal}
\author{C. E. Dahl} \affiliation{Department of Physics, Case Western Reserve University, Cleveland, OH 44106, USA}
\author{L. DeViveiros} \affiliation{Department of Physics, Brown University, Providence, RI 02912, USA}
\author{A.D. Ferella} \affiliation{Physics Institute, University of Z\"urich, Z\"urich, CH-8057, Switzerland}
\author{L.M.P. Fernandes} \affiliation{Department of Physics, University of Coimbra, Coimbra 3004-516, Portugal}
\author{S. Fiorucci} \affiliation{Department of Physics, Brown University, Providence, RI 02912, USA}
\author{R.J. Gaitskell} \affiliation{Department of Physics, Brown University, Providence, RI 02912, USA}
\author{K.L. Giboni} \affiliation{Department of Physics, Columbia University, New York, NY 10027, USA}
\author{R. Gomez} \affiliation{Department of Physics and Astronomy, Rice University, Houston, TX 77251, USA}
\author{R. Hasty} \affiliation{Department of Physics, Yale University, New Haven, CT 06511, USA}
\author{L. Kastens} \affiliation{Department of Physics, Yale University, New Haven, CT 06511, USA}
\author{J. Kwong} \affiliation{Department of Physics, Case Western Reserve University, Cleveland, OH 44106, USA}
\author{J.A.M. Lopes} \affiliation{Department of Physics, University of Coimbra, Coimbra 3004-516, Portugal}
\author{N. Madden} \affiliation{Lawrence Livermore National Laboratory, 7000 East Ave., Livermore, CA 94550, USA}
\author{A. Manalaysay} \affiliation{Physics Institute, University of Z\"urich, Z\"urich, CH-8057, Switzerland} 
\affiliation{Department of Physics, University of Florida, Gainesville, FL 32611, USA} 
\author{A. Manzur} \affiliation{Department of Physics, Yale University, New Haven, CT 06511, USA}
\author{D. N. McKinsey} \affiliation{Department of Physics, Yale University, New Haven, CT 06511, USA}
\author{M.E. Monzani} \affiliation{Department of Physics, Columbia University, New York, NY 10027, USA}
\author{K. Ni} \affiliation{Department of Physics, Yale University, New Haven, CT 06511, USA}
\author{U. Oberlack} \affiliation{Department of Physics and Astronomy, Rice University, Houston, TX 77251, USA}
\author{J. Orboeck} \affiliation{Department of Physics, RWTH Aachen University, Aachen, 52074, Germany}
\author{G. Plante} \affiliation{Department of Physics, Columbia University, New York, NY 10027, USA}
\author{R. Santorelli} \affiliation{Department of Physics, Columbia University, New York, NY 10027, USA}
\author{J.M.F. dos Santos} \affiliation{Department of Physics, University of Coimbra, Coimbra 3004-516, Portugal}
\author{P. Shagin} \affiliation{Department of Physics and Astronomy, Rice University, Houston, TX 77251, USA}
\author{T. Shutt} \affiliation{Department of Physics, Case Western Reserve University, Cleveland, OH 44106, USA}
\author{P. Sorensen} \affiliation{Department of Physics, Brown University, Providence, RI 02912, USA}
\author{S. Schulte} \affiliation{Department of Physics, RWTH Aachen University, Aachen, 52074, Germany}
\author{C. Winant} \affiliation{Lawrence Livermore National Laboratory, 7000 East Ave., Livermore, CA 94550, USA}
\author{M. Yamashita} \affiliation{Department of Physics, Columbia University, New York, NY 10027, USA}

\collaboration{XENON10 Collaboration}\noaffiliation

\date{\today}

\begin{abstract}
XENON10 is an experiment to directly detect weakly interacting massive particle (WIMPs),
 which may comprise the bulk of the non-baryonic dark matter in our Universe.  We report new  
 results for spin-dependent WIMP-nucleon interactions with $^{129}$Xe and $^{131}$Xe from 58.6 live-days of 
 operation at the Laboratori Nazionali del Gran Sasso (LNGS). Based on the non-observation of a WIMP signal in 5.4\,kg of 
 fiducial liquid xenon mass, we exclude previously unexplored regions in the theoretically allowed parameter space for neutralinos. 
 We also exclude a heavy Majorana neutrino with a mass in the range of $\sim$10\,GeV/c$^2$--2\,TeV/c$^2$ as a dark matter candidate under 
 standard assumptions for its density and distribution in the galactic halo.
\end{abstract}

\maketitle

Evidence for a significant cold dark matter component in our universe is stronger than ever \cite{freedman2003,bullet_cluster,dm_ring}, a well-motivated particle 
candidate being the lightest neutralino from super-symmetric extensions to the Standard Model \cite{jungman1996}.  Such a particle is neutral, non-relativistic, stable, 
and more generally 
classified as a Weakly Interacting Massive Particle (WIMP).  The open question of the nature of WIMPs is being addressed by numerous direct and indirect detection 
experiments \cite{gaitskell2004,gabriel05,laura06}.

Among these, the XENON10 experiment aims to directly detect galactic WIMPs scattering elastically from Xe atoms. Moving with velocities around 10$^{-3}c$, WIMPs can couple to 
nucleons via both spin-independent and spin-dependent (axial vector) interactions.  Spin-independent  
WIMP-nucleon couplings are in general smaller than axial vector couplings \cite{jungman1996}. However, 
for low momentum transfer,  they benefit from coherence across the nucleus, and therefore the overall event rate for WIMP interactions is 
expected to be dominated by the spin-independent coupling for target nuclei 
with A$\geq$30.  The sensitivity  of XENON10 to spin-independent interactions is published in \cite{Xe10_prl_SI}.

We report here on a spin-dependent analysis of 58.6 live-days of WIMP-search data, taken in low-background conditions at LNGS, 
which provides $\sim$3100 meters water equivalent rock overburden.  XENON10 is a dual phase (liquid and gas) xenon time projection 
chamber, discriminating between the predominantly electron-recoil background and the expected nuclear-recoil 
WIMP signal via the distinct ratio of ionization to scintillation for each type of interaction \cite{Xe10_prl_SI}.  A nuclear recoil energy 
threshold of 4.5\,keV was achieved, and 10 candidate events were recorded for an exposure of about 136\,kg\,days after analysis cuts (the fiducial mass was 5.4\,kg). 
Although all observed events are consistent with expected background from electron recoils (see \cite{Xe10_prl_SI} for details 
on the analysis and the candidate events),  no background subtraction is employed for calculating the WIMP upper limits.
In the following analysis, we use identical data quality, fiducial volume, and physics cuts as reported in  \cite{Xe10_prl_SI}.

For axial WIMP-nuclei interactions, the WIMPs couple to the spins of the nucleons. Although the interaction with the nucleus is coherent (as it is in the 
spin-independent case) in the sense that scattering amplitudes are summed over nucleons, the strength of the interaction vanishes for paired 
nucleons in the same energy state. Thus only nuclei with an odd number of nucleons will yield a significant sensitivity to axial WIMP-nuclei interactions 
\cite{lewin1996}. Almost half of naturally occurring xenon has non-zero nuclear spin: $^{129}$Xe (spin-1/2) makes up $26.4\%$, and $^{131}$Xe (spin-3/2) another $21.2\%$.  
The differential WIMP-nucleus cross section for the spin-dependent interaction can be written as \cite{jungman1996}:

\begin{eqnarray}
\frac{d\sigma}{d|\bf{q}|^2} = \frac{C_{spin}}{v^2}G_F^2\frac{ S(|\bf{q}|)}{S(0)} ,
\end{eqnarray}

\noindent
where $G_F$ is the Fermi constant, $v$ is the WIMP velocity relative to the target, $S(|\bf{q}|)$ is the spin structure 
function for momentum transfer $q>0$, and S(0) represents the zero-momentum transfer limit. The so called enhancement factor, $C_{spin}$, is given by

\begin{eqnarray}
C_{spin} = \frac{8}{\pi} [a_p\langle S_p\rangle + a_n\langle S_n\rangle ]^2  \frac{J+1}{J},
\end{eqnarray}

\noindent
where $J$ is the total nuclear spin, $a_p$ and $a_n$ are the effective WIMP-nucleon couplings (these depend on the quark spin distribution within the nucleons and 
on the WIMP type), and $\langle S_{p,n}\rangle=\langle N|S_{p,n}|N \rangle$ are the expectation values of the spin 
content of the proton and neutron groups within the nucleus.

Detailed nuclear shell-model calculations for $\langle S_{p,n}\rangle$ for two different Hamiltonians describing the nuclei exist in the literature \cite{ressell1997}.
The Hamiltonians are based on realistic nucleon-nucleon potentials,  Bonn A \cite{bonn} and Nijmegen II \cite{nijmegen}.
The accuracy of these calculations is typically assessed by comparing their predictions of the magnetic moment $\mu$ with experimental values,  
due to the similarity between the magnetic moment operator and the matrix element for WIMP-nucleon coupling.  

\begin{figure*}
\includegraphics[width=8.5cm]{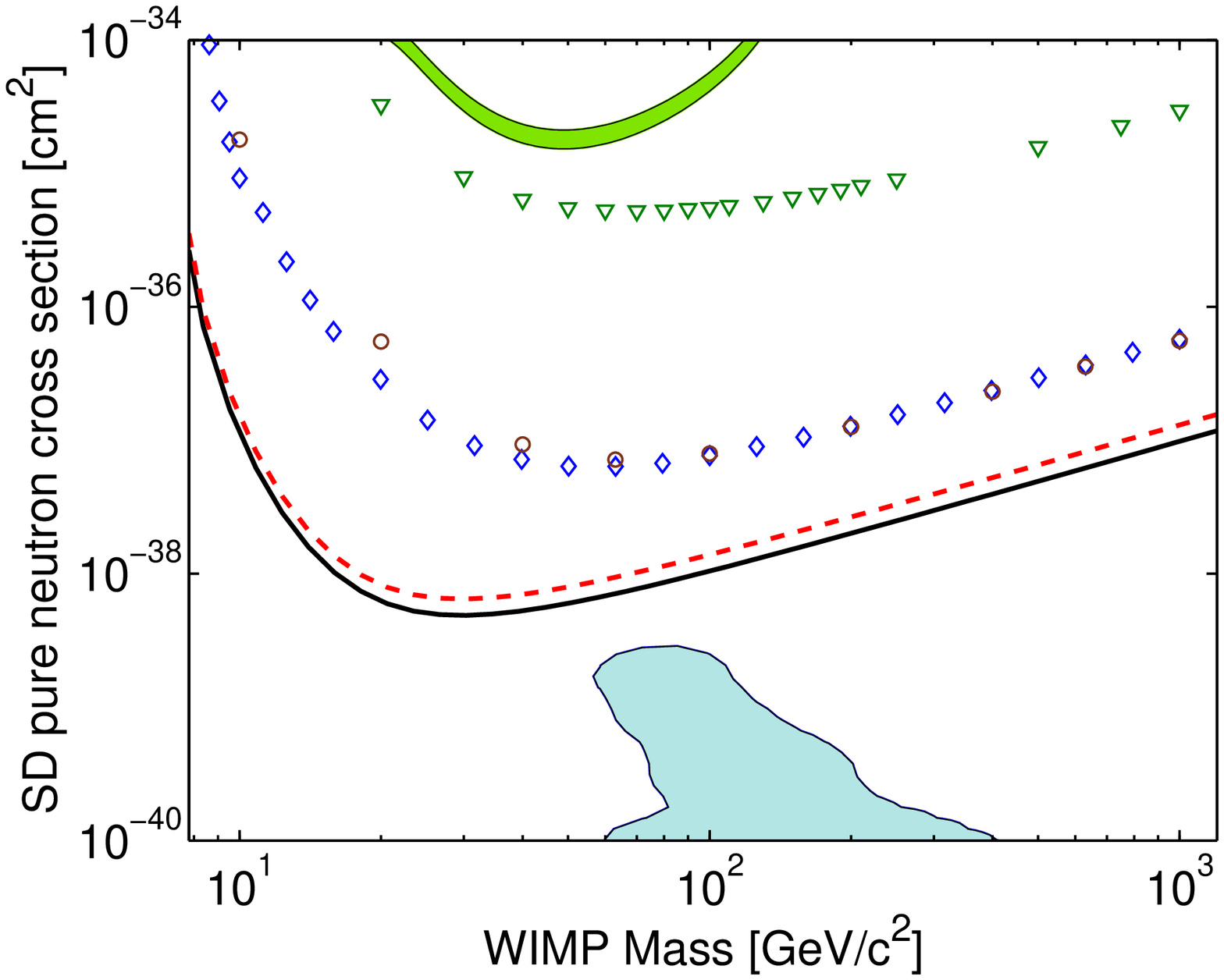}
\includegraphics[width=8.5cm]{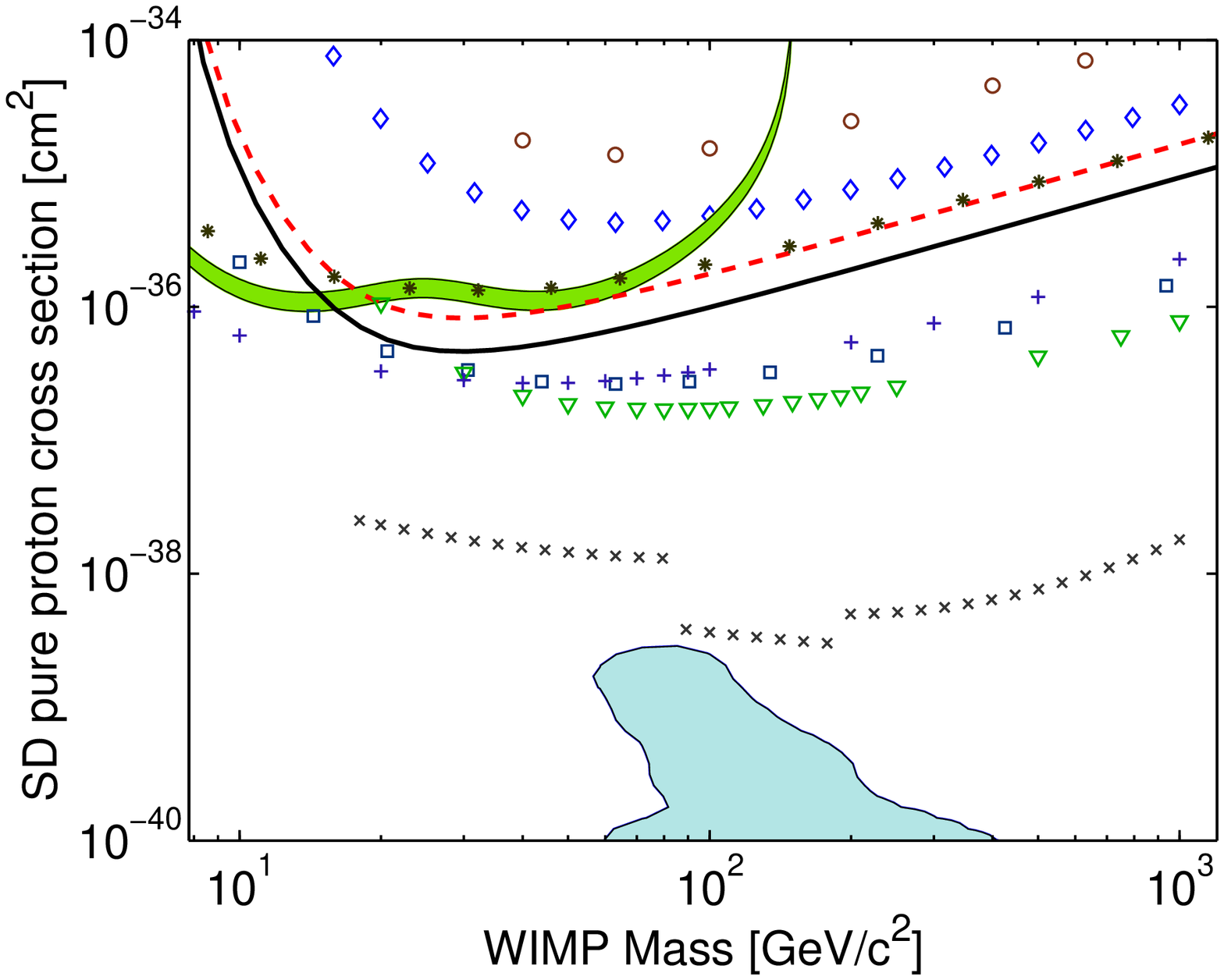}
\caption{XENON10 combined 90\% CL exclusion limits for $^{129}$Xe and $^{131}$Xe for pure neutron (left) and pure proton (right) couplings (solid curves). 
The dashed curves show the combined Xe limits using the alternate form factor. Also shown are the 
results from the CDMS experiment \cite{cdmsSD2006} (diamonds), ZEPLIN-II \cite{zep2} (circles), KIMS \cite{kims} (triangles), 
NAIAD \cite{naiad} (squares), PICASSO \cite{picasso} (stars), COUPP \cite{coupp} (pluses), SuperK \cite{superK2004} (crosses), as well 
as the DAMA evidence region under the assumption of standard WIMP nuclear recoils and dark halo parameters (green filled region) \cite{savage}. 
The theoretical regions (blue filled) for the neutralino (in the constrained minimal supersymmetric model) are taken from \cite{roberto2007}.}
\label{fig_1}
\end{figure*}

For $^{129}$Xe the magnetic moment agrees to within  25\% and 11\% for the Bonn A and Nijmegen II  potentials, respectively, using the standard free-particle g-factor   
(and to within 19\% and 52\% for using effective g-factors).  We calculate our main WIMP-nucleon exclusion limits using the Bonn A potential (Fig. \ref{fig_1}, solid curves), 
with $\langle{\bf S}_{p}\rangle =0.028$ and $\langle{\bf S}_{n}\rangle =0.359$.  In order to indicate the level of systematic uncertainty associated with the different models 
we also calculate limits using the alternate Nijmegen II potential (Fig. \ref{fig_1}, dashed curves), with  $\langle{\bf S}_{p}\rangle =0.0128$ and $\langle{\bf S}_{n}\rangle =0.300$.
For $^{131}$Xe, the same  Bonn A and Nijmegen II models predict the magnetic moment to within 8\% and 50\% of the measured value, respectively. However, for this isotope there are 
calculations by Engel using the Quasiparticle Tamm-Dancoff approximation (QTDA) \cite{engel1991}, which yield a magnetic moment within 1\% of the experimental value.  
We follow \cite{ressell1997,ressel_private} 
and use the calculation by Engel, choosing  $\langle{\bf S}_{p}\rangle =-0.041$ and $\langle{\bf S}_{n}\rangle = -0.236$, the effect of the 3 
different models for $^{131}$Xe on the variation in the exclusion limits being quite small.  We also note that recent calculations by Kortelainen, 
Toivanen and Toivanen \cite{kortelainen} yield values for the magnetic moments within about 20\% and 10\% for $^{129}$Xe and $^{131}$Xe, respectively, without using effective g-factors. 
For the sake of brevity and comparison with other experimental results (which follow \cite{ressell1997}), we use the $\langle S_{p,n}\rangle$ values detailed above.

In the limit of zero momentum transfer the WIMP essentially interacts with the entire nucleus. Once the momentum transfer $q$ reaches a magnitude such that 
$\hbar/q$ is no longer large compared to the nucleus, the spatial distribution of the nuclear spin must be considered.  
This is described by the spin structure function

\begin{eqnarray}
S(q) = a_0^2S_{00}(q) + a_0a_1S_{01}(q) + a_1^2S_{11}(q),
\end{eqnarray}

\noindent
here written by using the isospin convention in terms of the isoscalar ($a_0 = a_p+a_n$) and isovector ($a_1 = a_p-a_n$) coupling constants. 
The independent form factors, namely the pure isoscalar term $S_{00}$, the pure isovector term $S_{11}$ and the interference term $S_{01}$ can 
be obtained from detailed nuclear shell model calculations. Ressel and Dean \cite{ressell1997} parameterize the structure function in terms of 
$y=(qb/2\hbar)^{2}$, where  $b$ parameterizes the nuclear size with $b=\sqrt[6]{A}$\,fm $\simeq$\,2.3\,fm for heavy nuclei. 
For a WIMP mass of 100~GeV/c$^2$ with velocity 10$^{-3}$c, typical Xe nuclear recoil energies are below $\sim$25\,keV. We have restricted our search to 
below 26.9\,keV, resulting in a maximum momentum transfer of $q\simeq$\,81 MeV/c, i.e $\hbar/q\geq$\,2.4\,fm and y$<$0.25. 
For our analysis, we use the structure function calculated with the Bonn A and Nijmegen II potentials \cite{ressell1997} and 
with the QTDA method \cite{engel1991} for $^{129}$Xe and $^{131}$Xe, respectively.

In order to set limits on spin-dependent WIMP nucleon couplings, we 
follow the procedure described in \cite{tovey2000,giuliani2004,savage}, avoiding model-dependent assumptions 
on the (Majorana) WIMP composition.  
We first present exclusion limits for the cases of pure-proton ($a_n$ = 0) and pure-neutron ($a_p$ = 0) couplings, by assuming 
that the total cross section is dominated by the proton and neutron contributions only.
We calculate the 90\% C.L. exclusion limits as a function of WIMP mass with Yellin's Maximal Gap method \cite{yellin2002}.
The exclusion limits presented here assume a flat 19\% $\mathcal{L}_{eff}$ for the xenon scintillation efficiency of nuclear recoils 
relative to electron recoils.  The motivation for this choice is explained in reference \cite{Xe10_prl_SI}, in which we also show that 
the uncertainty in $\mathcal{L}_{eff}$ could raise the limits by about 15\% (18\%) for a WIMP mass of 30\,GeV/c$^2$ (100\,GeV/c$^2$).

In Fig.\ref{fig_1} we show the combined upper limit curves in the WIMP-nucleon cross section versus WIMP mass plane, for the 
simple case of pure neutron (left) and pure proton (right) couplings for the $^{129}$Xe and $^{131}$Xe isotopes.
We also include limits from CDMS \cite{cdmsSD2006}, ZEPLIN-II \cite{zep2}, KIMS \cite{kims}, 
NAIAD \cite{naiad}, PICASSO \cite{picasso}, COUPP \cite{coupp} as well as the indirect detection limits from Super-Kamiokande \cite{superK2004} 
for the case of pure proton couplings.
 Given that  $^{129}$Xe and $^{131}$Xe both contain an unpaired neutron, XENON10 is mostly sensitive to WIMP-neutron spin-dependent 
 couplings and excludes previously unexplored regions of parameter space. The minimum WIMP-nucleon cross section 
 of $\sim$6$\times$10$^{-39}$cm$^2$ is achieved at a WIMP mass of around 30\,GeV/c$^2$. 
 The sensitivity to pure proton-couplings is less strong, however XENON10 improves upon the parameter space constrained by the ZEPLIN-II \cite{zep2}
and the CDMS \cite{cdmsSD2006} experiments and approaches the 
 sensitivity of other direct detection experiments such NAIAD \cite{naiad}, PICASSO \cite{picasso}, KIMS\cite{kims} and COUPP \cite{coupp},  
 all containing nuclei with unpaired protons. 
 We also show the DAMA evidence region for standard WIMP interactions and halo parameters \cite{savage}, as well as  
 predictions of neutralino-nucleon cross sections in the constrained minimal super-symmetric model (CMSSM) \cite{roberto2007}.
 Although the expected cross sections are still below the current experimental spin-dependent sensitivity, direct WIMP detection experiments are now 
 for the first time approaching the theoretically predicted parameter space for neutralinos. 
 
 As a further benchmark, we consider heavy Majorana neutrinos with standard weak interactions. 
 Such neutrinos, with masses in the region 100-500\,GeV/c$^2$ have recently been proposed as dark matter 
 candidates in minimal technicolor theories in cosmologies with a dynamical dark energy term 
 \cite{kainulainen,kouvaris}. 
The expected cross section on protons and neutrons can be written as \cite{primack88}:
 
 \begin{eqnarray}
\sigma_{\nu N} = \frac{8G_F^2}{\pi\hbar^4}\mu^2 C_{spin,\nu} 
 \end{eqnarray}
 
 \noindent
 where $\mu$ is the neutrino-nucleon reduced mass, and the spin enhancement factor in this case is:

\begin{eqnarray}
C_{spin,\nu} =[a_p\langle S_p \rangle + a_n \langle S_n \rangle]^2 \frac{J+1}{J}
\end{eqnarray}

\noindent
 with the values for the WIMP-nucleon spin factors $a_{p}$ = 0.46 and 
 $a_{n}$ = 0.34 taking into account the strange quark contribution to the nucleon spin, 
 as measured by  the EMC collaboration and given in \cite{ellis95} for coupling to protons and 
 neutrons, respectively.

\begin{figure*}
\includegraphics[width=8.5cm]{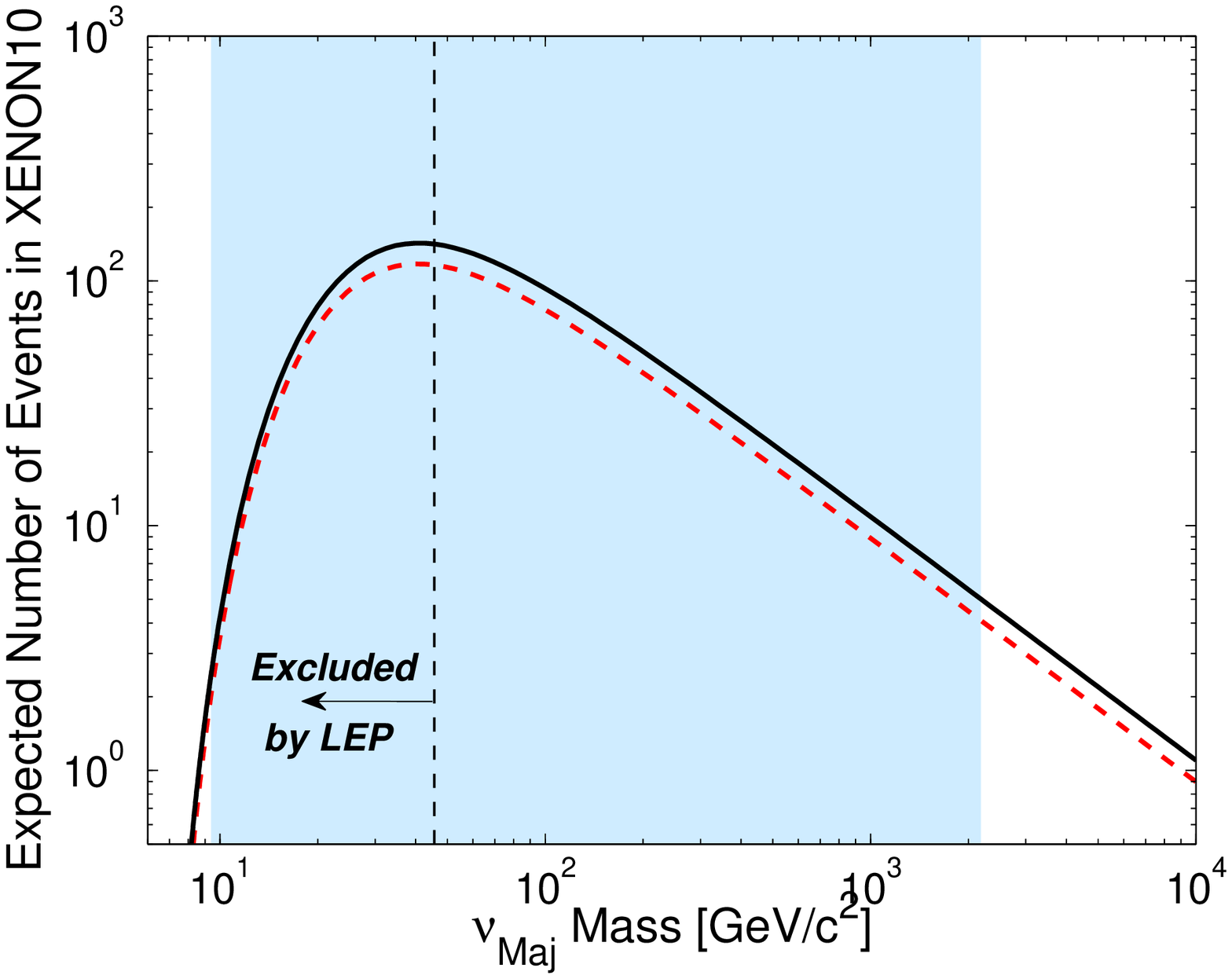}
\includegraphics[width=8.5cm]{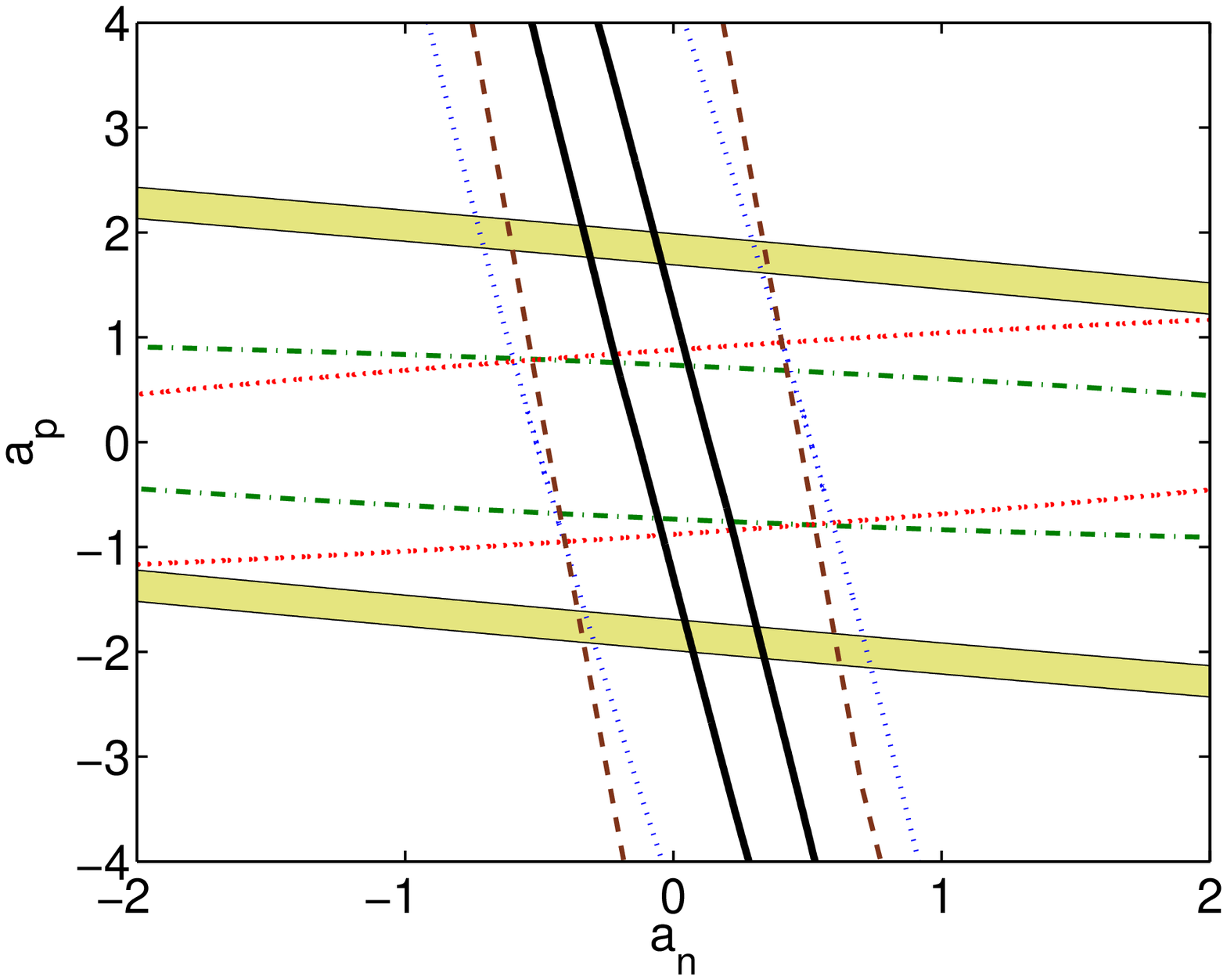}
\caption{Left: Predicted number of events in XENON10 for a heavy Majorana neutrino 
with standard weak interaction as a function of the neutrino mass, using the main (solid curve) and 
alternate (dashed curve) form factors. The light shaded area shows the excluded mass region at 90\% CL, 
calculated with Yellin's Maximal Gap method \cite{yellin2002} for the main form factors.
Right: Regions allowed at the 90\% CL 
in $a_n - a_p$ parameter space for a WIMP mass of 50\,GeV/c$^2$. The combined limit from $^{129}$Xe and $^{131}$Xe 
is shown as a dark solid curve (using the main form factor, see text), 
The exteriors of the corresponding ellipses are excluded, 
the common space inside the ellipses being allowed by the data. We also show the results obtained by KIMS \cite{kims} (dot-dashed), COUPP \cite{coupp} (dotted) ('horizontal' ellipses) and 
CDMS \cite{cdmsSD2006} (dotted), ZEPLIN-II \cite{zep2} (dashed) and the DAMA evidence region \cite{savage} (light filled region) ('vertical' ellipses).}
\label{fig2}
\end{figure*}

In Fig. \ref{fig2} (left) we show the predicted number of events in XENON10 as a function of the heavy  
Majorana neutrino mass, the light shaded area showing the excluded mass region at 90\% CL 
for using the main form factors. 
 Our result excludes a heavy Majorana 
neutrino as a dark matter candidate with a mass between 9.4\,GeV/c$^2$--2.2\,TeV/c$^2$ (9.6\,GeV/c$^2$--1.8\,TeV/c$^2$ 
for the alternate form factor). 
We note that a heavy Majorana neutrino with a mass below half the 
Z-boson mass has already been excluded at LEP \cite{lep}. 

We now present the results in terms of the more general phase space for $a_n$ and $a_p$ for a fixed WIMP mass. 
We follow \cite{savage} and express the expected number of recoil events N$_{xenon}$  as a function of $a_n$ and $a_p$:

\begin{eqnarray}
N_{xenon} = A a_p^2 + B a_p a_n + C a_n^2
\end{eqnarray}

\noindent
with A, B, C being constants of integration of the differential event rate $dR/dE$ over the relevant energy region, in our case 4.5\,keV--27\,keV 
nuclear recoil energy. 

Fig. \ref{fig2} (right) shows the allowed regions at 90\% CL in the  $a_p - a_n$ parameter space for a  
WIMP mass of 50\,GeV/c$^2$. We include the published CDMS \cite{cdmsSD2006}, ZEPLIN-II \cite{zep2},  KIMS \cite{kims}  and the DAMA allowed region \cite{savage}  
 for comparison. The advantage of using different isotopes with spin as dark matter targets is evident: 
the presence of both odd-neutron and odd-proton number isotopes  breaks the degeneracy and only the common space 
inside the ellipses is allowed by the data.

\noindent
In conclusion, we have obtained new limits on the spin-dependent WIMP-nucleon cross section by operating a 
liquid-gas xenon time projection chamber at  LNGS, in WIMP search mode for 58.6 live days with a fiducial mass of 5.4\,kg.
The results for pure neutron couplings are the world's most stringent to date, reaching a minimum cross section of 
5$\times$10$^{-39}$cm$^2$ at a WIMP mass of 30\,GeV/c$^2$. We exclude new regions in the  $a_p - a_n$ parameter space, 
and, for the first time, we directly probe a heavy Majorana neutrino as a dark matter candidate. Our observations exclude 
a heavy Majorana neutrino with a mass between $\sim$10\,GeV/c$^2$--2\,TeV/c$^2$ for using a local WIMP density of 0.3\,GeV/cm$^3$ and 
a Maxwell-Boltzmann velocity distribution. We note that our sensitivity to axial-vector couplings could be strongly improved by 
using a larger mass of enriched $^{129}$Xe as the dark matter target. 

This work was funded by NSF Grants No. PHY-03-02646 and No. PHY-04-00596, the CAREER Grant No. PHY-0542066, the DOE Grant No. DE-FG02-91ER40688, by the 
Swiss National Foundation SNF Grant No 20-118119, by the Volkswagen Foundation (Germany) and by the FCT Grant No POCI/FIS/60534/2004 (Portugal). 
We would like to thank the LNGS/INFN staff and engineers  for their help and support.



\end{document}